\documentstyle[12pt]{article}
\catcode`\@=11
\def\numberbysection{\@addtoreset{equation}{section}
\def\theequation{\thesection.\arabic{equation}}}

\numberbysection
\newcommand{\frad}[2]{\displaystyle{\displaystyle#1\over\displaystyle#2}}
\newcommand\fd{fluc\-tu\-a\-tion-dis\-si\-pa\-tion }
\begin{document}
\centerline{\large\bf Correlation and response in the Backgammon model:}
\vspace{.3cm}
\centerline{\large\bf the Ehrenfest legacy}
\vspace{1cm}
\centerline{\large
by C.~Godr\`eche$^{a,}$\footnote{godreche@spec.saclay.cea.fr}
and J.M.~Luck$^{b,}$\footnote{luck@spht.saclay.cea.fr}}
\vspace{1cm}
\centerline{$^a$Service de Physique de l'\'Etat Condens\'e,
CEA Saclay, 91191 Gif-sur-Yvette cedex, France}
\vspace{.1cm}
\centerline{$^b$Service de Physique Th\'eorique,
CEA Saclay, 91191 Gif-sur-Yvette cedex, France}
\vspace{1cm}
\begin{abstract}
We pursue our investigation of the non-equilibrium dynamics
of the Backgammon model, a dynamical urn model
which exhibits aging and glassy behavior at low temperature.
We present an analytical study of the scaling behavior
of the local correlation and response functions of the density fluctuations
of the model, and of the associated \fd ratios,
throughout the $\alpha$-regime of low temperature and long times.
This analysis includes the aging regime, the convergence to equilibrium,
and the crossover behavior between them.
\end{abstract}
\vfill
\vskip -6pt
\noindent P.A.C.S.~02.50.Ey, 05.40.+j, 61.43.Fs, 75.50.Lk
\newpage
\section{Introduction}
\setcounter{footnote}{0}

There is a long tradition in Statistical Physics to illustrate conceptual
problems on extremely simplified models~\cite{ehr1}.
The Ehrenfest dynamical urn model~\cite{ehr2},
which was considered by Kac as `probably one of the
most instructive models in the whole of Physics'~\cite{kac1},
is an example of these.
It was devised to elucidate the problems raised by the objections
of Loschmidt and Zermelo against the $H$-theorem, and in particular
to understand the fine details of how a system relaxes towards equilibrium.

The Ehrenfest model is defined as follows.
Consider $N$ balls, labeled from $1$ to $N$, distributed between two urns.
The numbers of balls in the urns are denoted by $N_1$ and $N_2=N-N_1$.
At each time step a ball is chosen at random
(i.e., an integer between 1 and $N$ is chosen at random),
and moved from the box where it is to the other box.
If there was initially an unequal number
of balls in each urn, the system will relax to the equilibrium state,
characterized by the binomial law
\begin{equation}
P(N_1)=\frac{1}{2^N}{N\choose N_1},
\label{bine}
\end{equation}
such that each urn contains, on average, an equal number of balls.
This model, subsequently studied by Kohlrausch and
Schr\"{o}dinger~\cite{schro}, was finally fully solved by Kac,
Siegert and Hess~\cite{kac2,sieg,hess}.

The present paper belongs to the same tradition.
It is devoted to the analytical study of a very simple urn model,
which is actually a modern avatar of the Ehrenfest model,
the so-called Backgammon model~\cite{ritort}.
This model pertains to the recent stream of research where the study of the
slow dynamics of complex glassy systems is replaced by that of simple
models, most of them of mean-field type~\cite{revue}.
Its aim was to
demonstrate that a microscopic model characterized by the absence of energy
barriers was nevertheless able to exhibit a number of the characteristic
features of glassy dynamics.

The Backgammon model generalizes the Ehrenfest model in two directions.
First, it is easy to realize that the dynamics of the Ehrenfest model
is effectively at infinite temperature,
because there is no constraint on the move of the drawn ball.
Therefore, in order to consider the dynamics of the model
at finite temperature, one has to define an energy and impose e.g.
a Metropolis
rule for the move of the drawn ball: the move is allowed with probability
one if the energy decreases or stays unchanged, and with probability
$\exp(-\beta\Delta E)$ if it increases by $\Delta E$.
The second generalization consists in taking $M$ boxes instead of two,
and in going to the thermodynamical limit: $N\to\infty$, $M\to\infty$,
with a fixed density $\rho=N/M$ of balls per box.
These two generalizations define a class of `thermodynamical urn models',
of which the Backgammon model is a special case.
In the Backgammon model the energy is
chosen equal to minus the number of empty boxes~\cite{ritort}.

Variants of this model have been considered, where either the choice of
energy is different, or the a priori statistics of the elementary moves
is defined otherwise~\cite{gbm,kim,bial,dgc}, or yet other rules are
imposed~\cite{prados}.
Mentions of the similarity of the Backgammon model to the
Ehrenfest model can also be found in refs.~\cite{lipo,kehr}.

The expected behavior of the model is easy to grasp.
Denoting by $N_{d}$ the occupation number of the departure box,
determined by the selected ball,
and by $N_{a}$ the occupation number of the arrival box, drawn at random,
four elementary moves are possible:
\begin{eqnarray}
&&\left.
\begin{array}{l}
N_{d}>1,\,N_{a}>0:\,\Delta E=0\\
N_{d}=1,\,N_{a}=0:\,\Delta E=0
\end{array}
\right\}\;{\rm (diffusion),}\nonumber\\
&&
\begin{array}{l}
N_{d}=1,\,N_{a}>0:\,\Delta E=-1
\end{array}
\;{\rm (condensation),}\nonumber\\
&&
\begin{array}{l}
N_{d}>1,\,N_{a}=0:\,\Delta E=+1
\end{array}
\;{\rm (evaporation).}
\label{elem}
\end{eqnarray}
The first two moves, corresponding to diffusion of the balls, are
independent of temperature.
The last two ones are respectively favored
at low temperature (condensation), and at high temperature
(evaporation).\footnote{We use on purpose
the same words --diffusion, condensation and
evaporation-- as used in the description of the elementary moves in the
Kawasaki dynamics of an Ising chain~\cite{corn}.
There is indeed a strong analogy between the latter model
and a one-dimensional version of the Backgammon model.}

Therefore, at infinite temperature, since once a ball is drawn it is replaced
in a box chosen at random, equilibrium is attained rapidly, characterized by
a multinomial distribution of the balls amongst the boxes, a simple
generalization of the result for the original Ehrenfest model.
When temperature decreases, dynamics
slows down because the rate of rejection of moves of the drawn ball increases.
At equilibrium and at low temperature, a few boxes contain a large number of
balls.
Finally at zero temperature the only
moves allowed are those for which energy decreases (condensation) or stays
unchanged (diffusion).
The dynamics becomes extremely slow as time passes.
Indeed the ultimate stage when all boxes but two are empty
meets the original Ehrenfest model with two boxes, except that now,
if one box becomes empty, the dynamics stops.
Such an event occurs with a probability of order
$P(N)=2^{-N}$ [see eq.~(\ref{bine})],
so that the time needed to empty one of the two boxes
in this ultimate stage is exponentially large in $N$,
and infinite in the thermodynamical limit.
It is therefore clear that, at low temperature, after a transient regime,
two time scales are involved in the process:
a fast one, corresponding to the equilibration amongst the non-empty boxes,
and a slow one, corresponding to the occurrence of the rare
event of emptying a new box.
This picture, though simple, accounts for the gist of the model.

A virtue of the Backgammon model is that it allows to test a number of
concepts which appear in the study of the non-equilibrium dynamics of real
systems, though it is an extreme simplification of the latter.
For instance,
slow relaxational dynamics, aging properties, exploration of the two-time
plane, violation of the \fd theorem~\cite{ck}, etc.
can be investigated in this model, either by numerical means~\cite{ritort},
in a straightforward way, or by analytical methods, which
turn out to be much harder~\cite{fr1,gbm,fr2,gl96,fr3,gl97}.

The relaxation of energy for the Backgammon model was the subject of
previous studies~\cite{fr1,gbm,fr2,gl96,gl97}.
More recently the
relaxation of density has been addressed by Franz and Ritort~\cite{fr3},
in order to investigate possible differences between the behavior of this
observable and that of the energy.
In the present work we give a thorough analytical study of this question.
The methods used are similar to those introduced in ref.~\cite{gl97}.
Our presentation will therefore follow the lines of this reference,
yet stay self-contained.

The outline of this paper is as follows.
Sec.~2 defines the quantities
studied in this work and presents the dynamical equations they obey.
In Sec.~3 a formal solution of these equations is obtained by the
method of characteristics.
Sec.~4 makes contact with the original Ehrenfest
model by considering the simple situation of infinite temperature.
The two main sections of this work are Sec.~5, devoted to equilibrium
properties,
and especially Sec.~6, which contains a detailed investigation of the
non-equilibrium $\alpha $-regime at low temperature.
Sec.~7 gives a brief summary and a discussion.

\section{Physical quantities and their dynamical evolution}

Consider a finite system, made of $M$ boxes containing $N$ particles.
Let $N_{i}(t)$ be the occupation number of box number $i$ at time $t$,
i.e., the number of particles contained in that box.
We have
\begin{equation}
\sum_{i=1}^{M}N_{i}(t)=N.
\end{equation}
The Hamiltonian ${\cal H}$ and action $S$ of the system at inverse
temperature $\beta$ read
\begin{equation}
S\equiv\beta{\cal H}=-\beta\sum_{i=1}^{M}\delta_{N_{i},0}.
\end{equation}

Instead of addressing the more ambitious goal of describing the dynamics of
the model by the configuration ${\cal C}=\left\{N_{1},\ldots, N_{M}\right\} $
involving all the occupation numbers, we restrict our study to
quantities involving the occupation number of one box only.
We consider the thermodynamical limit ($N\to\infty$, $M\to\infty$),
with a fixed density $\rho=N/M$, and
furthermore restrict the analysis to the homogeneous non-equilibrium initial
condition where there is one particle per box.
We thus have $N=M$, $\rho=1$, and $N_{i}(t=0)=1$ for $i=1,\dots,M$.

\subsection{Reminder: occupation probabilities and mean energy}

Taking box number 1 as a generic box, we denote by $f_{k}(t)$ the
probability that it contains $k$ particles at time $t$:
\begin{equation}
f_{k}(t)={\rm Prob}\left\{N_{1}(t)=k\right\}.
\end{equation}

The dynamical equations obeyed by the occupation probabilities $f_{k}(t)$
read [see e.g. Appendix~A of ref.~\cite{gl97} for a derivation]
\begin{eqnarray}
\frac{{\rm d}f_{k}(t)}{{\rm d}t} &=&\frac{k+1}{\Lambda(t)}
f_{k+1}(t)+f_{k-1}(t)-\left(1+{{\frac{k}{\Lambda(t)}}}\right)
f_{k}(t)\qquad(k\ge 2){,}\nonumber\\
\frac{{\rm d}f_{1}(t)}{{\rm d}t} &=&\frac{2}{\Lambda(t)}f_{2}(t)+\mu
(t)f_{0}(t)-2f_{1}(t),\nonumber\\
\frac{{\rm d}f_{0}(t)}{{\rm d}t} &=&f_{1}(t)-\mu(t)f_{0}(t),
\label{eqfk}
\end{eqnarray}
with the initial value
\begin{equation}
f_{k}(0)=\delta_{k,1},
\label{init1}
\end{equation}
and where
\begin{equation}
{{\frac{1}{\Lambda(t)}}}=1+({\rm e}^{-\beta}-1)f_{0}(t),\qquad
\mu(t)={\rm e}^{-\beta}+(1-{\rm e}^{-\beta})f_{1}(t).
\label{lamudef}
\end{equation}

Eqs.~(\ref{eqfk}) describe a birth-death process,
i.e., a random walk on the $k$-axis with position-dependent rates.
The rate at which a ball enters a non-empty box ($k\ge1$) is equal to 1,
while it is equal to $\mu$ for an empty box.
The first process is temperature-independent as it should, since diffusion
between non-empty boxes is not constrained by temperature.
The second process depends on temperature.
For instance, at infinite temperature, $\mu$ is equal to 1,
because moving a ball to an empty box
is always allowed (diffusion and evaporation are both permitted).
At zero temperature, $\mu$ is equal to $f_{1}$, which expresses the fact that
moving a ball to an empty box is allowed if the departure box contains one
ball only (only diffusion is permitted).
The departure rate of a ball from a box containing $k$ balls is equal to
$k/\Lambda(t)$.
Hence, there is equilibration of $k$ around $\Lambda(t)$.
At low temperature, $\Lambda(t)\approx(1-f_{0}(t))^{-1}$,
which is equal to the average number of balls in a non-empty box.

To summarize, eqs.~(\ref{eqfk}) describe a random walk along the $k$-axis,
in a confining potential centered around $\Lambda(t)$, which itself is
slowly increasing, whenever the number of empty boxes increases~\cite{gbm}.
It is indeed intuitively clear that while the equilibration inside the
potential is fast,
the evolution of $\Lambda(t)$ is very slow at low temperature.
This defines two time scales, as mentioned in the Introduction, or two regimes,
which are to be identified respectively with the $\beta$- and the
$\alpha$-regime, characteristic of glassy relaxation~\cite{revue}.

Eqs.~(\ref{eqfk}) can be formally rewritten as a Markovian system, of the form
\begin{equation}
\frac{{\rm d}f_{k}(t)}{{\rm d}t}=\sum_{\ell\ge 0}{\cal M}_{k\ell
}[f_{0}(t),f_{1}(t)]f_{\ell}(t),\label{compact}
\end{equation}
where the matrix ${\cal M}[f_{0}(t),f_{1}(t)]$ satisfies
\begin{equation}
\sum_{k\ge 0}{\cal M}_{k\ell}[f_{0}(t),f_{1}(t)]=0.
\label{conserv}
\end{equation}
Actually, the reaction terms $\Lambda(t)$ and $\mu(t)$ involve
$f_{0}(t)$ and $f_{1}(t)$, which makes the dynamics
non-Markovian and the system of equations~(\ref{eqfk}) or~(\ref{compact})
non-linear.

Eqs.~(\ref{eqfk}) or~(\ref{compact}) ensure the conservation of the moments
\begin{equation}
\sum_{k\ge 0}f_{k}(t)=1,\qquad\langle N_{1}(t)\rangle=\sum_{k\ge
1}k\,f_{k}(t)=1,
\label{eqn1}
\end{equation}
respectively corresponding to the total numbers of boxes and of particles.
Finally, the mean energy per box of the system reads
$E(t)=-\langle\delta_{N_1(t),0}\rangle=-f_{0}(t).$

\subsection{Density correlation function}

The density correlation function $c(t,s)$ measures the correlation between
the number of particles in box number 1 at times $s$ (waiting time) and $t$
(observation time), with $0\le s\le t$:
\begin{equation}
c(t,s)=\langle N_{1}(t)N_{1}(s)\rangle -\langle N_{1}(t)\rangle\langle
N_{1}(s)\rangle=\langle N_{1}(t)N_{1}(s)\rangle -1.
\label{cdef}
\end{equation}
By definition, we have
\begin{equation}
\langle N_{1}(t)N_{1}(s)\rangle=\sum_{j,k\ge 1}jk\,{\rm Prob}\left\{
N_{1}(t)=k|N_{1}(s)=j\right\}{\rm Prob}\left\{N_{1}(s)=j\right\}.
\end{equation}
Eq.~(\ref{cdef}) can thus be rewritten as
\begin{equation}
c(t,s)=\sum_{j,k\ge 1}jkf_{j}(s)g_{k}^{(j)}(t,s)-1,
\label{cts}
\end{equation}
where
\begin{equation}
g_{k}^{(j)}(t,s)={\rm Prob}\left\{N_{1}(t)=k|N_{1}(s)=j\right\},
\end{equation}
or else as
\begin{equation}
c(t,s)=\sum_{k\ge 1}k\gamma_{k}(t,s)-1,
\end{equation}
with
\begin{equation}
\gamma_{k}(t,s)=\sum_{j\ge 1}jf_{j}(s)g_{k}^{(j)}(t,s).
\end{equation}
Since the time variable $s$ plays the role of a parameter in the dynamics,
both $g_{k}^{(j)}(t,s)$ and $\gamma_{k}(t,s)$ obey the dynamical equations
\begin{eqnarray}
{{\frac{\partial g_{k}^{(j)}(t,s)}{\partial t}}} &=&\sum_{\ell\ge 0}
{\cal M}_{k\ell}[f_{0}(t),f_{1}(t)]g_{\ell}^{(j)}(t,s),
\label{eqgk}\\
{{\frac{\partial\gamma_{k}(t,s)}{\partial t}}} &=&\sum_{\ell\ge 0}
{\cal M}_{k\ell}[f_{0}(t),f_{1}(t)]\gamma_{\ell}(t,s),
\label{eqgamk}
\end{eqnarray}
with the initial conditions
\begin{equation}
g_{k}^{(j)}(s,s)=\delta_{k,j},\qquad\gamma_{k}(s,s)=kf_{k}(s),
\label{init2}
\end{equation}
implying in particular
\begin{equation}
c(s,s)=\langle N_1(s)^2\rangle-1=\sum_{k\ge 1}k^2\,f_k(s)-1,
\end{equation}
in agreement with eq.~(\ref{cdef}).
Eqs.~(\ref{eqgk}, \ref{eqgamk}) ensure the conservation of probability, i.e.,
\begin{equation}
\sum_{k\ge 0}g_{k}^{(j)}(t,s)=1,\qquad
\sum_{k\ge 0}\gamma_{k}(t,s)=1.
\label{consgam}
\end{equation}

\subsection{Density response functions}

The local density response function $r(t,s)$ is a measure of the change in
the density of box number 1 at time $t$, induced by an infinitesimal change
in the conjugate variable, i.e., in the present case the local chemical
potential acting on the same box, at the earlier time $s$.
Assume that box number 1 is subjected to an arbitrary
time-dependent chemical potential, given by $\alpha(t)$ in reduced units.
The perturbed action now reads
\begin{equation}
S(t)=-\beta\sum_{i=1}^{M}\delta_{N_{i}(t),0}-\alpha(t)N_{1}(t).
\end{equation}

The occupation probabilities of box number 1 now depend on $\alpha(t)$:
we denote them by $f_{k}^{\alpha}(t)={\rm Prob}\left\{N_1(t)=k\right\}$.
In the thermodynamical limit, i.e., to leading order as $M\to\infty$,
the occupation probabilities of all
the other boxes $(i=2,\ldots,M)$ are still given by the $f_{k}(t)$.
Indeed the mean-field geometry of the model implies that the mutual influence
of any two distinct boxes scales as $1/M$.

Assuming for simplicity $|\alpha(t)|<\beta$, the $f_{k}^{\alpha}(t)$ obey
the dynamical equations
\begin{eqnarray}
\frac{{\rm d}f_{k}^{\alpha}(t)}{{\rm d}t} &=&(k+1)\nu
_{-}(t)f_{k+1}^{\alpha}(t)+\nu_{+}(t)f_{k-1}^{\alpha}(t)-(\nu
_{+}(t)+k\nu_{-}(t))f_{k}^{\alpha}(t)\qquad(k\ge 2),\nonumber\\
\frac{{\rm d}f_{1}^{\alpha}(t)}{{\rm d}t} &=&2\nu_{-}(t)f_{2}^{\alpha
}(t)+\mu_{+}(t)f_{0}^{\alpha}(t)-(\nu_{+}(t)+\mu_{-}(t))f_{1}^{\alpha
}(t),\nonumber\\
\frac{{\rm d}f_{0}^{\alpha}(t)}{{\rm d}t} &=&\mu_{-}(t)f_{1}^{\alpha
}(t)-\mu_{+}(t)f_{0}^{\alpha}(t),
\label{eqfalk}
\end{eqnarray}
with the initial value
\begin{equation}
f_{k}^{\alpha}(0)=\delta_{k,1},
\end{equation}
and with the definitions
\begin{eqnarray}
\mu_{+}(t) &=&f_{1}(t){\cal W}(-\alpha(t))+(1-f_{1}(t)){\rm e}^{-\beta
+\alpha(t)},\nonumber\\
\mu_{-}(t) &=&1-f_{0}(t)+f_{0}(t){\cal W}(\alpha(t)),\nonumber\\
\nu_{+}(t) &=&f_{1}(t)+(1-f_{1}(t)){\cal W}(-\alpha(t)),\nonumber\\
\nu_{-}(t)&=&(1-f_{0}(t)){\cal W}(\alpha(t))+f_0(t){\rm e}^{-\beta-\alpha(t)}.
\end{eqnarray}
These equations can be derived along the same lines as in ref.~\cite{gl97}.

In the above expressions,
\begin{equation}
{\cal W}(\Delta S)=\min(1,{\rm e}^{-\Delta S})
\end{equation}
is the Metropolis acceptance rate associated with a change of action
$\Delta S\equiv\beta\Delta E$.
This rate is not differentiable at $\Delta S=0$.
We have indeed
\begin{equation}
\left.\frac{{\rm d}{\cal W}}{{\rm d}\Delta S}\right|_{\Delta S\to0^+}=-1,
\qquad
\left.\frac{{\rm d}{\cal W}}{{\rm d}\Delta S}\right|_{\Delta S\to0^-}=0.
\label{metrodis}
\end{equation}
As a consequence, we are led to define the following two local
response functions for the density
\begin{equation}
r^{\pm}(t,s)=\left.{{\frac{\delta\langle N_{1}(t)\rangle}{\delta\alpha
(s)}}}\right|_{\alpha(s)\to 0^{\pm}}.
\label{r+-}
\end{equation}
Introducing the quantities
\begin{equation}
h_{k}^{\pm}(t,s)
=\left.{{\frac{\delta f_{k}^{\alpha}(t)}{\delta\alpha(s)}}}\right|_
{\alpha(s)\to 0^{\pm}},
\end{equation}
we have
\begin{equation}
r^{\pm}(t,s)=\sum_{k\ge 1}kh_{k}^{\pm}(t,s).
\end{equation}

The dynamical equations for the $h_{k}^{\pm}(t,s)$ are obtained by taking
the functional derivative of eqs.~(\ref{eqfalk}) with respect to $\alpha(s)$
for $\alpha(s)\to 0^{\pm}$.
This yields
\begin{equation}
\frac{\partial h_{k}^{\pm}(t,s)}{\partial t}=\sum_{\ell\ge 0}
{\cal M}_{k\ell}[f_{0}(t),f_{1}(t)]h_{\ell}^{\pm}(t,s),
\label{eqhk}
\end{equation}
with the initial conditions
\begin{eqnarray}
h_{k}^{+}(s,s) &=&-\frac{k+1}{\Lambda(s)}f_{k+1}(s)+\frac{k}{\Lambda(s)}
f_{k}(s)\qquad(k\ge 2),\nonumber\\
h_{1}^{+}(s,s) &=&-\frac{2}{\Lambda(s)}f_{2}(s)+\mu(s)f_{0}(s),\nonumber\\
h_{0}^{+}(s,s) &=&-\mu(s)f_{0}(s),\nonumber\\
h_{k}^{-}(s,s) &=&-(k+1){\rm e}^{-\beta
}f_{0}(s)f_{k+1}(s)\nonumber\\
&&+(1-f_{1}(s))f_{k-1}(s)
+(k{\rm e}^{-\beta}f_{0}(s)+f_{1}(s)-1)f_{k}(s)\qquad(k\ge 2),\nonumber\\
h_{1}^{-}(s,s) &=&-2{\rm e}^{-\beta}f_{0}(s)f_{2}(s)+\mu
(s)f_{0}(s)+(f_{1}(s)-1)f_{1}(s),\nonumber\\
h_{0}^{-}(s,s) &=&-\mu(s)f_{0}(s).
\label{hpm}
\end{eqnarray}
We have
\begin{equation}
\sum_{k\ge 0}h_{k}^{\pm}(t,s)=0
\label{consh}
\end{equation}
at all times $t\ge s$.
Furthermore, eqs.~(\ref{hpm}) imply that
both response functions coincide at initial times:
\begin{equation}
r^{+}(s,s)=r^{-}(s,s)=1+(2{\rm e}^{-\beta}-1)f_{0}(s)-f_{1}(s)
+2(1-{\rm e}^{-\beta})f_{0}(s)f_{1}(s).
\end{equation}

\subsection{Fluctuation-dissipation ratios}

The \fd ratio provides a measure of the violation of the \fd theorem,
and thus of the departure of the system from equilibrium~\cite{revue,ck,fr3}.
In the present case, we need to define two \fd ratios
\begin{equation}
X^{\pm}(t,s)=\frac{r^{\pm}(t,s)}{\,\frad{\partial c(t,s)}{\partial s}\,}.
\end{equation}
These definitions contain no explicit temperature dependence, because a
factor of temperature has been absorbed in the definition of the reduced
chemical potential $\alpha(t)$.

We are thus led to investigate the quantity
\begin{equation}
\frac{\partial c(t,s)}{\partial s}=\sum_{k\ge 1}k\zeta_{k}(t,s),
\end{equation}
with
\begin{equation}
\zeta_{k}(t,s)={{\frac{\partial\gamma_{k}(t,s)}{\partial s}}}.
\end{equation}
For $t>s$, the dynamical equations obeyed by these quantities
are deduced from eq.~(\ref{eqgamk}) and read
\begin{equation}
{{\frac{\partial\zeta_{k}(t,s)}{\partial t}}}=\sum_{\ell\ge 0}
{\cal M}_{k\ell}[f_{0}(t),f_{1}(t)]\zeta_{\ell}(t,s).
\label{eqdzk}
\end{equation}
Eq.~(\ref{hpm}) implies
\begin{equation}
\sum_{k\ge 0}\zeta_{k}(t,s)={{\frac{\partial}{\partial s}}}\sum_{k\ge
0}\gamma_{k}(t,s)=0.
\label{consdz}
\end{equation}
Moreover, by integrating eq.~(\ref{eqgamk}) for the $\gamma_{k}(t,s)$ to
first order in $\theta=t-s$, we obtain the initial values
\begin{equation}
\zeta_{k}(s,s)=\sum_{\ell\ge 0}(k-\ell){\cal M}_{k\ell
}[f_{0}(t),f_{1}(t)]f_{\ell}(t),
\end{equation}
i.e.,
\begin{eqnarray}
\zeta_{k}(s,s) &=&-\frac{k+1}{\Lambda(s)}f_{k+1}(s)+f_{k-1}(s)\qquad(k\ge
2),\nonumber\\
\zeta_{1}(s,s) &=&-\frac{2}{\Lambda(s)}f_{2}(s)+\mu(s)f_{0}(s),\nonumber\\
\zeta_{0}(s,s) &=&-f_{1}(s),
\label{init3}
\end{eqnarray}
implying
\begin{equation}
\left.\frac{\partial c(t,s)}{\partial s}\right|_{t=s}=2-\frac{c(s,s)}
{\Lambda(s)}+f_{0}(s)\left(\mu(s)-1\right).
\end{equation}

\section{Generating functions and integral representations}

It is possible to obtain integral representations of the solutions of
eqs.~(\ref{eqfk}, \ref{eqgamk}, \ref{eqhk}, \ref{eqdzk}) by means of the
generating functions
\begin{eqnarray}
F(x,t) &=&\sum_{k\ge 0}f_{k}(t)\,x^{k},\qquad G(x,t,s)=\sum_{k\ge 0}\gamma
_{k}(t,s)\,x^{k},\nonumber\\
H^{\pm}(x,t,s) &=&\sum_{k\ge 0}h_{k}^{\pm}(t,s)\,x^{k},\qquad
Z(x,t,s)=\sum_{k\ge 0}\zeta_{k}(t,s)\,x^{k}.
\end{eqnarray}
These functions obey the partial differential equations
\begin{eqnarray}
\frac{\partial F(x,t)}{\partial t} &=&(x-1)\left(F(x,t)-{{\frac{1}{\Lambda
(t)}\frac{\partial F(x,t)}{\partial x}}}-Y_{f}(t)\right),\nonumber\\
\frac{\partial G(x,t,s)}{\partial t} &=&(x-1)\left(G(x,t,s)-{{\frac{1}
{\Lambda(t)}\frac{\partial G(x,t,s)}{\partial x}}}-Y_{\gamma}(t,s)\right),
\nonumber\\
\frac{\partial H^{\pm}(x,t,s)}{\partial t} &=&(x-1)\left(H^{\pm}(x,t,s)
-{{\frac{1}{\Lambda(t)}\frac{\partial H^{\pm}(x,t,s)}{\partial x}}}
-Y_{h^{\pm}}(t,s)\right),\nonumber\\
\frac{\partial Z(x,t,s)}{\partial t} &=&(x-1)\left(Z(x,t,s)-{{\frac{1}
{\Lambda(t)}\frac{\partial Z(x,t,s)}{\partial x}}}-Y_{\zeta}(t,s)\right).
\label{partialad}
\end{eqnarray}
The $Y$-functions which enter these equations read
\begin{eqnarray}
Y_{f}(t) &=&(1-{\rm e}^{-\beta})f_{0}(t)=1-1/\Lambda(t),\nonumber\\
Y_{\gamma}(t,s) &=&(1-{\rm e}^{-\beta})\left(\gamma_{0}(t,s)+f_{0}(t)
{{\frac{\partial\gamma_{0}(t,s)}{\partial t}}}-{{\frac{{\rm d}f_{0}(t)}
{{\rm d}t}}}\gamma_{0}(t,s)\right){,}\nonumber\\
Y_{h^{\pm}}(t,s) &=&(1-{\rm e}^{-\beta})\left(h_{0}^{\pm}(t,s)+f_{0}(t)
{{\frac{\partial h_{0}^{\pm}(t,s)}{\partial t}}}-{{\frac{{\rm d}f_{0}(t)}
{{\rm d}t}}}h_{0}^{\pm}(t,s)\right){,}\nonumber\\
Y_{_{\zeta}}(t,s) &=&(1-{\rm e}^{-\beta})\left(\zeta_{0}(t,s)+f_{0}(t)
{{\frac{\partial\zeta_{0}(t,s)}{\partial t}}}-{{\frac{{\rm d}f_{0}(t)}
{{\rm d}t}}}\zeta_{0}(t,s)\right){.}
\label{ydzeta}
\end{eqnarray}

The initial conditions for eq.~(\ref{partialad}) are derived from
eqs.~(\ref{init1}, \ref{init2}, \ref{hpm}, \ref{init3}):
\begin{eqnarray}
F(x,0) &=&x,\nonumber\\
G(x,s,s) &=&x\,\frac{\partial F(x,s)}{\partial x},\nonumber\\
H^{+}(x,s,s) &=&(x-1)\left({{\frac{1}{\Lambda(s)}\frac{\partial F(x,s)}
{\partial x}}}+{\rm e}^{-\beta}f_{0}(s)-f_{1}(s)+2(1-{\rm e}^{-\beta
})f_{0}(s)f_{1}(s)\right){,}\nonumber\\
H^{-}(x,s,s) &=&(x-1)\nonumber\\
&\times&\left((1-f_{1}(s))F(x,s)+{\rm e}^{-\beta}f_{0}(s)\frac
{\partial F(x,s)}{\partial x}
+(1-{\rm e}^{-\beta})f_{0}(s)(2f_{1}(s)-1)\right),\nonumber\\
Z(x,s,s) &=&-\frac{1}{\Lambda(s)}\frac{\partial F(x,s)}{\partial x}
+xF(x,s)-(1-{\rm e}^{-\beta})f_{0}(s)\left((1-x)f_1(s)+x\right).
\label{zss}
\end{eqnarray}
The conservation of moments expressed by eqs.~(\ref{eqn1}, \ref{consgam},
\ref{consh}, \ref{consdz}) implies
\begin{equation}
F(1,t)=\left.{{\frac{\partial F(x,t)}{\partial x}}}\right|
_{x=1}=G(1,t,s)=1,\qquad H^{\pm}(1,t,s)=Z(1,t,s)=0.
\end{equation}

Eqs.~(\ref{partialad}) can be solved by the method of characteristics
[see e.g. Appendix~B of ref.~\cite{gl97}].
We thus obtain
\begin{eqnarray}
F(x,t) &=&\left(1+(x-1){\rm e}^{-\tau(t)}\right)
{\rm e}^{(x-1)D(t,0)}+\int_{0}^{t}{\rm d}u\,{\cal K}(x,t,u)Y_{f}(u),
\nonumber\\
G(x,t,s) &=&{\rm e}^{(x-1)D(t,s)}\,G\!\left(1+(x-1){\rm e}^{\tau(s)-\tau
(t)},s,s\right) +\int_{s}^{t}{\rm d}u\,{\cal K}(x,t,u)Y_{\gamma}(u,s),
\label{eqFG}
\end{eqnarray}
and expressions for $H^{\pm}(x,t,s)$ and $Z(x,t,s)$ similar to that for
$G(x,t,s)$, with the definitions
\begin{eqnarray}
\tau(t) &=&\int_{0}^{t}\frac{{\rm d}u}{\Lambda(u)},\\
D(t,u) &=&\int_{u}^{t}{\rm d}v\,{\rm e}^{\tau(v)-\tau(t)},\\
{\cal K}(x,t,u) &=&(1-x){\rm e}^{\tau(u)-\tau(t)+(x-1)D(t,u)}.
\end{eqnarray}

The quantities of interest are given by
\begin{equation}
c(t,s)=\left.{{\frac{\partial G(x,t,s)}{\partial x}}}\right|_{x=1}-1,
\ r^{\pm}(t,s)=\left.{{\frac{\partial H^{\pm}(x,t,s)}{\partial x}}}\right|
_{x=1},\ \frac{\partial c(t,s)}{\partial s}=\left.{{\frac{\partial Z(x,t,s)}
{\partial x}}}\right|_{x=1}.
\label{eqint}
\end{equation}
Using eqs.~(\ref{eqFG}) and the corresponding equations for $H^{\pm}(x,t,s)$
and $Z(x,t,s)$, we obtain finally
\begin{eqnarray}
c(t,s) &=&D(t,s)-1+(c(s,s)+1){\rm e}^{\tau(s)-\tau(t)}-\int_{s}^{t}{\rm d}
u\,{\rm e}^{\tau(u)-\tau(t)}Y_{\gamma}(u,s){,}\nonumber\\
r^{\pm}(t,s) &=&r^{\pm}(s,s)\,{\rm e}^{\tau(s)-\tau(t)}-\int_{s}^{t}{\rm d}
u\,{\rm e}^{\tau(u)-\tau(t)}Y_{h^{\pm}}(u,s),\nonumber\\
\frac{\partial c(t,s)}{\partial s} &=&\frac{\partial c(s,s)}{\partial s}
\,{\rm e}^{\tau(s)-\tau(t)}-\int_{s}^{t}{\rm d}u\,{\rm e}^{\tau(u)-\tau
(t)}Y_{\zeta}(u,s).
\label{dcts}
\end{eqnarray}
Setting $x=0$ in eqs.~(\ref{eqFG}), we get integral representations
of the following quantities, which will be useful in Sec.~6:
\begin{eqnarray}
f_{0}(t) &=&(1-{\rm e}^{-\tau(t)}){\rm e}^{-D(t,0)}+\int_{0}^{t}{\rm d}u\,
{\cal K}(0,t,u)Y_{f}(u),\nonumber\\
\gamma_{0}(t,s) &=&{\rm e}^{-D(t,s)}\,G\!\left(1-{\rm e}^{\tau(s)-\tau
(t)},s,s\right) +\int_{s}^{t}{\rm d}u\,{\cal K}(0,t,u)Y_{\gamma}(u,s),
\label{eqgam0}
\end{eqnarray}
with
\begin{equation}
{\cal K}(0,t,u)={\rm e}^{\tau(u)-\tau(t)-D(t,u)}={{\frac{{\partial}}
{{\partial}u}}}{\rm e}^{-D(t,u)},
\end{equation}
and similarly
\begin{eqnarray}
h_{0}^{\pm}(t,s) &=&{\rm e}^{-D(t,s)}H^{\pm}\left(1-{\rm e}^{\tau
(s)-\tau(t)},s,s\right) +\int_{s}^{t}{\rm d}u\,{\cal K}(0,t,u)Y_{_{h^{\pm
}}}(u,s),\nonumber\\
\zeta_{0}(t,s) &=&{\rm e}^{-D(t,s)}Z\left(1-{\rm e}^{\tau(s)-\tau
(t)},s,s\right) +\int_{s}^{t}{\rm d}u\,{\cal K}(0,t,u)Y_{\zeta}(u,s).
\label{eqdzeta0}
\end{eqnarray}

\section{Infinite-temperature behavior}

At infinite temperature, the Backgammon model,
which is a `thermodynamical' generalization of the Ehrenfest model,
as discussed in the Introduction, is solvable.
The solutions given in eqs.~(\ref{eqFG}, \ref{dcts}) indeed become explicit,
because of the simplifications ${\rm e}^{-\beta}=\Lambda(t)=\mu(t)=1$.

\subsection{Reminder: occupation probabilities and mean energy}

Since $Y_{f}(t)=0$, we have
\begin{equation}
F(x,t)=(1+(x-1){\rm e}^{-t}){\rm e}^{(x-1)(1-{\rm e}^{-t})}.
\end{equation}
The occupation probabilities thus read
\begin{equation}
f_{k}(t)=((1-{\rm e}^{-t})^{2}+k{\rm e}^{-t}){{\frac{(1-{\rm e}^{-t})^{k-1}
{\rm e}^{{\rm e}^{-t}-1}}{k!}}},
\end{equation}
yielding, in particular, the mean energy
\begin{equation}
E(t)=-f_{0}(t)=-(1-{\rm e}^{-t}){\rm e}^{{\rm e}^{-t}-1}.
\end{equation}

\subsection{Correlation and response, \fd ratios}

Similarly, noticing that $Y_{\gamma}(t,s)=Y_{h^{\pm}}(t,s)=Y_{\zeta}(t,s)=0$,
we get
\begin{eqnarray}
G(x,t,s) &=&\left(1+(x-1){\rm e}^{s-t}\right)\left(1+(x-1){\rm e}^{-t}
(1-{\rm e}^{-s})\right){\rm e}^{(x-1)(1-{\rm e}^{-t})}{,}\nonumber\\
H^{+}(x,t,s) &=&(x-1)\nonumber\\
&\times&\left[\left({{\rm e}^{s}+(x-1){\rm e}^{-t}
({\rm e}^{-s}-1)}\right){\rm e}^{(x-1)(1-{\rm e}^{-t})-t}
-{\rm e}^{(x-1)(1-{\rm e}^{s-t})-s-t+{\rm e}^{-s}-1}\right],\nonumber\\
H^{-}(x,t,s) &=&(x-1){\rm e}^{(x-1)(1-{\rm e}^{-t})-t}\nonumber\\
&\times&\left[\left({1+(x-1){\rm e}^{-t}}\right)
{\rm e}^{(x-1)s}-\left({1+(x-1){\rm e}^{s-t}}\right)
{\rm e}^{-s+{\rm e}^{-s}-1}\right],\nonumber\\
Z(x,t,s) &=&(x-1)\left({(x-1){\rm e}^{-t}+1+{\rm e}^{-2s}}\right)
{\rm e}^{(x-1)(1-{\rm e}^{-t})+s-t},
\end{eqnarray}
We thus have
\begin{equation}
c(t,s)=(1-{\rm e}^{-2s}){\rm e}^{s-t},\quad r^{\pm}(t,s)=(1-{\rm e}^{-2s+
{\rm e}^{-s}-1}){\rm e}^{s-t},\quad{{\frac{\partial c(t,s)}{\partial s}}}
=(1+{\rm e}^{-2s}){\rm e}^{s-t}.
\end{equation}
Note that both response functions coincide for all times at infinite
temperature: $r^{+}(t,s)=r^{-}(t,s)$, whereas the generating functions
$H^{\pm}(x,t,s)$ are different from each other.
The common value of the \fd ratios is independent of $t$, and reads
\begin{equation}
X^{\pm}(t,s)=\frac{1-{\rm e}^{-2s+{\rm e}^{-s}-1}}{1+{\rm e}^{-2s}}.
\label{eqxts}
\end{equation}
At equilibrium, i.e., for $s\gg 1$, the correlation and response functions
only depend on the time difference $\theta=t-s$, and they fulfil the
\fd theorem (see Sec.~5):
\begin{equation}
c_{{\rm eq}}(\theta)=r_{{\rm eq}}^{\pm}(\theta)
=-\frac{{\rm d}c_{{\rm eq}}(\theta)}{{\rm d}\theta}={\rm e}^{-\theta}.
\end{equation}
Equivalently, $X^{\pm}(t,s)$ in eq.~(\ref{eqxts})
goes to unity as $s\to\infty$.

\section{Equilibrium properties}

\subsection{Reminder: occupation probabilities and mean energy}

At any finite temperature and for long enough times, the system converges
towards thermal equilibrium.
The occupation probabilities at equilibrium are
given by the stationary solution of the dynamical equations~(\ref{eqfk}):
\begin{eqnarray}
(f_{k})_{{\rm eq}} &=&{\rm e}^{-\Lambda_{{\rm eq}}}
\frac{\Lambda_{{\rm eq}}^{k-1}}{k!}\qquad(k\ge 1),\nonumber\\
-E_{{\rm eq}} &=&(f_{0})_{{\rm eq}}=\frac{\Lambda_{{\rm eq}}-1
+{\rm e}^{-\Lambda_{{\rm eq}}}}{\Lambda_{{\rm eq}}}
=\frac{{\rm e}^{\beta-\Lambda_{{\rm eq}}}}{\Lambda_{{\rm eq}}}{,}
\end{eqnarray}
where $\Lambda_{{\rm eq}}$, the equilibrium value of $\Lambda(t)$, is related
to temperature by
\begin{equation}
{\rm e}^{\beta}=1+(\Lambda_{{\rm eq}}-1){\rm e}^{\Lambda_{{\rm eq}}}.
\end{equation}
This quantity can be identified with the thermodynamical fugacity of the
model [see Appendix~C of ref.~\cite{gl97}].
Note that the most probable
non-zero occupancy is approximately equal to $\Lambda_{{\rm eq}}$,
in agreement with the discussion given below eq.~(\ref{lamudef}).

At low temperature, we have
\begin{eqnarray}
{\Lambda_{{\rm eq}}} &=&{\beta-\ln\beta+{{\frac{\ln\beta+1}{\beta}}}
+{{\frac{\ln ^{2}\beta-1}{2\beta^{2}}}}+\cdots ,}\nonumber\\
E_{{\rm eq}} &=&-1+\frac{1}{\beta}+\frac{\ln\beta}{\beta^{2}}+\frac{\ln
^{2}\beta-\ln\beta-1}{\beta^{3}}{+\cdots}
\end{eqnarray}

Finally, the generating function of the equilibrium occupation probabilities
reads
\begin{equation}
F_{{\rm eq}}(x)={{\frac{\Lambda_{{\rm eq}}-1+
{\rm e}^{(x-1)\Lambda_{{\rm eq}}}}{\Lambda_{{\rm eq}}}}}.
\label{feqx}
\end{equation}

The convergence towards equilibrium is characterized by a relaxation time
$\tilde{t}_{{\rm eq}}$ (denoted as $t_{\rm eq}^{(1)}$ in ref.~\cite{gl97}),
which is exponentially divergent at low temperature~[see eq.~(\ref{tun})].

\subsection{Density correlation and response functions}

At equilibrium, the density correlation function $c(t,s)=c_{{\rm eq}}(\theta)$
and response functions $r^\pm(t,s)=r^\pm_{{\rm eq}}(\theta)$ are
stationary: they only depend on the time difference $\theta=t-s$.

These quantities can be derived from eqs.~(\ref{dcts}),
where all integrals become convolutions.
These equations can therefore be solved in Laplace space.
Denoting by $\hat{c}_{{\rm eq}}(p)$
the Laplace transform of $c_{{\rm eq}}(\theta)$, and using similar
conventions for the other functions, we obtain after some algebra
\begin{equation}
\hat{r}_{{\rm eq}}^{+}(p)=\hat{r}_{{\rm eq}}^{-}(p)=\Lambda_{{\rm eq}}
-p\hat{c}_{{\rm eq}}(p),
\end{equation}
which express that both density response functions
coincide at equilibrium, and that they obey the identity
\begin{equation}
r_{{\rm eq}}^{\pm}(\theta)=-\frac{{\rm d}c_{{\rm eq}}(\theta)}
{{\rm d}\theta}.
\end{equation}
In other words, the \fd ratios $X^{\pm}(\theta)$
are identically equal to unity at equilibrium, expressing thus the \fd theorem.

Furthermore, we obtain the explicit results
\begin{eqnarray}
\hat{r}_{{\rm eq}}^\pm(p)=\frac{\Lambda_{{\rm eq}}-(\Lambda_{{\rm eq}}-1)
\left(p+1/(f_{0})_{{\rm eq}}\right)
\left(\hat{K}(p,\Lambda_{{\rm eq}})-{{\frad{\Lambda_{{\rm eq}}
{\rm e}^{-\Lambda_{{\rm eq}}}}{1+p\Lambda_{{\rm eq}}}}}\right)}
{(1+p\Lambda_{{\rm eq}})\left(\Lambda_{{\rm eq}}-(\Lambda_{{\rm eq}
}-1)\left(p+1/(f_{0})_{{\rm eq}}\right)\hat{K}(p,\Lambda_{{\rm eq}})\right)},
\label{ligne2}
\end{eqnarray}
where
\begin{equation}
\hat{K}(p,\Lambda_{{\rm eq}})=\Lambda_{{\rm eq}}\int_{0}^{1}{\rm d}
z\,z^{p\Lambda_{{\rm eq}}}\,{\rm e}^{\Lambda_{{\rm eq}}(z-1)}={\rm e}
^{-\Lambda_{{\rm eq}}}\sum_{k\ge 0}\frac{1}{p+k/\Lambda_{{\rm eq}}}\,
\frac{\Lambda_{{\rm eq}}^{k-1}}{(k-1)!}
\end{equation}
is the Laplace transform of the equilibrium kernel
\begin{equation}
K(\theta ,\Lambda_{{\rm eq}})=\exp\left(-\theta /\Lambda_{{\rm eq}}
-\Lambda_{{\rm eq}}\left(1-{\rm e}^{-\theta/\Lambda_{{\rm eq}}}\right)\right).
\end{equation}

The expression~(\ref{ligne2}) is a meromorphic function, whose denominator
coincides with that found in ref.~\cite{gl97}
in the case of the energy correlation and response functions.
This denominator has an infinite
sequence of zeros on the real negative axis, which we denote by
$p=-p_{k}^{(2)}$, with $k\ge 1$.
The relaxation time of the equilibrium correlation and response functions
$t_{{\rm eq}}$ (denoted as $t_{\rm eq}^{(2)}$ in ref.~\cite{gl97}) is given by
the inverse of the smallest one:
\begin{equation}
t_{{\rm eq}}={{\frac{1}{p_{1}^{(2)}}}}.
\end{equation}
At low temperature, $t_{{\rm eq}}$ is exponentially divergent
($\alpha$-relaxation),
while all the other characteristic times remain microscopic,
i.e., of order $\Lambda_{{\rm eq}}$ ($\beta$-relaxation).
Indeed, we have
\begin{eqnarray}
{t_{{\rm eq}}} &\approx &\frac{(\Lambda_{{\rm eq}}-1){\rm e}^{\Lambda
_{{\rm eq}}}}{\Lambda_{{\rm eq}}^{2}}\left({\Lambda_{{\rm eq}}^{2}
{\rm e}^{-\Lambda_{{\rm eq}}}
I(\Lambda_{{\rm eq}})+1-\Lambda_{{\rm eq}}}\right)\nonumber\\
&\approx &\frac{2{\rm e}^{\Lambda_{{\rm eq}}}}{\Lambda_{{\rm eq}}}
\left(1+{{\frac{2}{\Lambda_{{\rm eq}}^{2}}}}+\cdots\right)
\approx\frac{2{\rm e}^{\beta}}{\beta^{2}}
\left(1+{{\frac{2\ln\beta+1}{\beta}}}+\cdots\right),
\label{teq}
\end{eqnarray}
with
\begin{equation}
I(\Lambda)=\int_{0}^{1}\frac{{\rm d}z}{z}({\rm e}^{\Lambda z}-1)=\sum_{n\ge
1}\frac{\Lambda ^{n}}{n\,n!}\approx\frac{{\rm e}^{\Lambda}}{\Lambda}
\sum_{\ell\ge 0}\frac{\ell !}{\Lambda ^{\ell}}.
\end{equation}

The density correlation and response functions read
\begin{equation}
r_{{\rm eq}}^{\pm}(\theta)=\sum_{k\ge 1}A_{k}{\rm e}^{-p_{k}^{(2)}\theta
},\qquad c_{{\rm eq}}(\theta)=\sum_{k\ge 1}\frac{A_{k}}{p_{k}^{(2)}}
{\rm e}^{-p_{k}^{(2)}\theta},
\label{rceq}
\end{equation}
where the $A_{k}$ are the residues of the expression~(\ref{ligne2})
at the poles $p=-p_{k}^{(2)}$.
In particular the values of these functions at equal times
$(\theta=0)$ provide the following sum rules for the residues $A_{k}$:
\begin{equation}
r_{{\rm eq}}^{\pm}(0)=\sum_{k\ge 1}A_{k}=\frac{1+(\Lambda_{{\rm eq}}-1)
{\rm e}^{-\Lambda_{{\rm eq}}}}{\Lambda_{{\rm eq}}},\qquad c_{{\rm eq}}(0)
=\sum_{k\ge 1}\frac{A_{k}}{p_{k}^{(2)}}=\Lambda_{{\rm eq}}.
\end{equation}

The only difference between the expressions~(\ref{rceq}) of the density
correlation
and response functions at equilibrium and those for the energy~\cite{gl97}
lies in the values of the residues: $A_k$ for the density
and $a_k$ for the energy, with
\begin{equation}
\frac{A_{k}}{a_{k}}=\left(\frac{(\Lambda_{{\rm eq}}-1)\left(
p_{k}^{(2)}-1/(f_{0})_{{\rm eq}}\right)}{\Lambda_{{\rm eq}}p_{k}^{(2)}-1}
\right)^2.
\end{equation}

Finally, the analysis presented in this section leads to the following
description of the behavior of the equilibrium density
correlation and response functions at low temperature.

\begin{itemize}
\item{In the $\beta$-regime $(\theta\sim 1)$, expression~(\ref{ligne2})
simplifies to
\begin{equation}
\hat{r}_{{\rm eq}}^{\pm}(p)=\Lambda_{{\rm eq}}-p\hat{c}_{{\rm eq}}(p)
\approx\frac{1}{1+p\Lambda_{{\rm eq}}},
\end{equation}
up to an exponentially small correction.
We thus obtain
\begin{equation}
r_{{\rm eq}}^{\pm}(\theta)\approx{{\frac{{\rm e}^{-\theta /\Lambda
_{{\rm eq}}}}{\Lambda_{{\rm eq}}}}},\qquad c_{{\rm eq}}(\theta)\approx
\Lambda_{{\rm eq}}-1+{\rm e}^{-\theta /\Lambda_{{\rm eq}}}.
\end{equation}
}
\item{In the $\alpha$-regime $(\theta\sim t_{{\rm eq}})$,
the expressions~(\ref{rceq}) are dominated by the first term $(k=1)$.
The corresponding residue can be estimated as
$A_{1}\approx(\Lambda_{{\rm eq}}-1)/t_{{\rm eq}}$.
We thus have
\begin{equation}
r_{{\rm eq}}^{\pm}(\theta)\approx{{\frac{\Lambda_{{\rm eq}}-1}
{t_{{\rm eq}}}}}{\rm e}^{-\theta /t_{{\rm eq}}},\qquad
c_{{\rm eq}}(\theta)\approx
(\Lambda_{{\rm eq}}-1){\rm e}^{-\theta /t_{{\rm eq}}}.
\label{ceq}
\end{equation}
}
\item{In the crossover between both regimes $(1\ll\theta\ll t_{{\rm eq}})$,
the density correlation and response functions exhibit
the plateau values
\begin{equation}
(r_{{\rm eq}}^{\pm})_{{\rm pl}}\approx\frac{\Lambda_{{\rm eq}}-1}
{t_{{\rm eq}}},\qquad(c_{{\rm eq}})_{{\rm pl}}\approx\Lambda_{{\rm eq}}-1.
\label{eqplateau}
\end{equation}
}
\end{itemize}

\section{Non-equilibrium behavior at low temperature}

In this section, we shall extend to density fluctuations
the low-temperature analysis developed in refs.~\cite{gl96,gl97}.
This analytical approach is valid at low temperature,
throughout the non-equilibrium $\alpha$-regime ($s\gg1$, $t-s\gg1$),
irrespective of the relative values of $s$ or $t$ with respect to
the relaxation time $t_{{\rm eq}}$.

\subsection{Reminder: mean energy}

In the case of the mean energy, our approach consists in
simplifying the first integral equation of eq.~(\ref{eqgam0}),
by expanding the integrand as a power series in $t-u$.
Keeping consistently the first two terms of this expansion,
we are left with the following differential equation
\begin{equation}
f_{0}(t)\approx{\rm e}^{-\Lambda(t)}+J_{0}(t)Y_{f}(t)+J_{1}(t)\frac{{\rm d}
Y_{f}(t)}{{\rm d}t},
\end{equation}
with
\begin{eqnarray}
J_{0}(t)&=&\int_{0}^{\infty}{\rm d}\varepsilon\,K(\varepsilon
,\Lambda(t))=\hat{K}(0,\Lambda(t))=1-{\rm e}^{-\Lambda(t)},\nonumber\\
J_{1}(t)&=&-\int_{0}^{\infty}{\rm d}\varepsilon\,\varepsilon
\,K(\varepsilon,\Lambda(t))=\left.{{\frac{{\rm d}\hat{K}(p,\Lambda(t))}
{{\rm d}p}}}\right|_{p=0}=-\Lambda(t){\rm e}^{-\Lambda(t)}I(\Lambda(t)).
\end{eqnarray}

This approach leads to the evolution equation~\cite{gl97}
\begin{equation}
\frac{{\rm d}\Lambda}{{\rm d}t}\approx A(\Lambda(t),\Lambda_{{\rm eq}}),
\label{dlambda}
\end{equation}
with
\begin{equation}
A(\Lambda ,\Lambda_{{\rm eq}})=\frac{1}{I(\Lambda)}\left(1-{{\frac
{(\Lambda -1){\rm e}^{\Lambda}}{(\Lambda_{{\rm eq}}-1)
{\rm e}^{\Lambda_{{\rm eq}}}}}}\right).
\end{equation}

Eq.~(\ref{dlambda}) describes the relaxation of energy throughout the
$\alpha$-regime at low temperature.
In particular,

\begin{itemize}
\item{At zero temperature, and more generally in
the aging regime $(1\ll t\ll t_{{\rm eq}})$, we have
\begin{equation}
\frac{{\rm d}\Lambda}{{\rm d}t}\approx\frac{1}{I(\Lambda)},
\end{equation}
hence
\begin{equation}
t\approx\int_{0}^{\Lambda}{\rm d}\Lambda ^{\prime}\,I(\Lambda ^{\prime
})=\int_{0}^{1}\frac{{\rm d}z}{z^{2}}({\rm e}^{\Lambda z}-1-\Lambda
z)=\sum_{n\ge 1}\frac{\Lambda ^{n+1}}{n(n+1)!}\approx\frac{{\rm e}^{\Lambda
}}{\Lambda}\sum_{\ell\ge 0}\frac{(\ell +1)!}{\Lambda ^{\ell}},
\end{equation}
or else
\begin{equation}
\Lambda(t)\approx\ln t+\ln\ln t+\frac{\ln\ln t-2}{\ln t}+\cdots
\end{equation}
}
\item{
In the opposite limit $(t\gg t_{{\rm eq}})$, the convergence of
$\Lambda(t)$ towards its equilibrium value $\Lambda_{{\rm eq}}$ is exponential,
\begin{equation}
\Lambda_{{\rm eq}}-\Lambda(t)\sim{\rm e}^{-t/\tilde t_{{\rm eq}}},
\end{equation}
with a relaxation time
\begin{eqnarray}
\tilde t_{{\rm eq}} &\approx &\frac{\Lambda_{{\rm eq}}-1}{\Lambda_{{\rm eq}}}
I(\Lambda_{{\rm eq}})\nonumber\\
&\approx &{{{\frac{{\rm e}^{\Lambda_{{\rm eq}}}}{\Lambda_{{\rm eq}}}}}
\left(1+{{\frac{1}{\Lambda_{{\rm eq}}^{2}}}}+\cdots\right)\approx
{{\frac{{\rm e}^{\beta}}{\beta^{2}}}}\left(1+{{\frac{2\ln\beta+1}
{\beta}}}+\cdots\right)}
\label{tun}
\end{eqnarray}
which is roughly equal to half the relaxation time $t_{{\rm eq}}$
of the correlation and response functions, given in eq.~(\ref{teq}).
}
\end{itemize}

\subsection{Correlation and response, \fd ratios}

The low-temperature approach consists in simplifying the integral
equations~(\ref{eqgam0}, \ref{eqdzeta0}) for $\gamma_{0}(t,s)$,
$h_{0}^{\pm}(t,s)$, and $\zeta_{0}(t,s)$ into the differential equations
\begin{eqnarray}
\gamma_{0}(t,s) &\approx &{\rm e}^{-\Lambda(t)}+J_{0}(t,s)Y_{\gamma
}(t,s)+J_{1}(t,s)\frac{\partial Y_{\gamma}(t,s)}{\partial t}{,}\nonumber\\
h_{0}^{\pm}(t,s) &\approx &J_{0}(t,s)Y_{h^{\pm}}(t,s)+J_{1}(t,s)
\frac{\partial Y_{h^{\pm}}(t,s)}{\partial t},\nonumber\\
\zeta_{0}(t,s) &\approx &J_{0}(t,s)Y_{\zeta}(t,s)+J_{1}(t,s)\frac{\partial
Y_{\zeta}(t,s)}{\partial t}.
\end{eqnarray}
These expressions lead, after some algebra, to the differential evolution
equation
\begin{equation}
\frac{1}{\phi(t,s)}\frac{{\partial}\phi(t,s)}{{\partial}t}\approx
-B(\Lambda(t),\Lambda_{{\rm eq}}),
\label{evolt}
\end{equation}
where $\phi(t,s)$ denotes either $\gamma_{0}(t,s)-f_{0}(t)$,
$h_{0}^{\pm}(t,s)$, or $\zeta_{0}(t,s)$, and with
\begin{equation}
B(\Lambda ,\Lambda_{{\rm eq}})=\frac{A(\Lambda ,\Lambda_{{\rm eq}})
+\Lambda ^{2}\left({\rm e}^{-\Lambda}
+{{\frad{{\rm e}^{-\Lambda_{{\rm eq}}}}
{\Lambda_{{\rm eq}}-1}}}\right)}{\Lambda
\left(\Lambda ^{2}{\rm e}^{-\Lambda}I(\Lambda)+1-\Lambda\right)}.
\end{equation}
Changing time variables from $s$ and $t$ to $\Lambda(s)$ and $\Lambda(t)$,
we obtain the alternative form
\begin{equation}
{{\frac{1}{\phi(t,s)}\frac{{\partial}\phi(t,s)}{{\partial}\Lambda(t)}}}
\approx -\alpha(\Lambda(t),\Lambda_{{\rm eq}}),
\label{evoll}
\end{equation}
with
\begin{equation}
\alpha(\Lambda ,\Lambda_{{\rm eq}})=\frac{B(\Lambda ,\Lambda_{{\rm eq}})}
{A(\Lambda ,\Lambda_{{\rm eq}})}.
\end{equation}
The normalized solution of the differential equation~(\ref{evoll}) reads
\begin{equation}
\phi(t,s)\approx\exp\left(-\int_{\Lambda(s)}^{\Lambda(t)}{\rm d}
\Lambda ^{\prime}\,\alpha(\Lambda ^{\prime},\Lambda_{{\rm eq}})\right).
\end{equation}
This result can be recast into a multiplicative scaling law
\begin{equation}
\phi(t,s)\approx{{\frac{\Phi(\Lambda(s),\Lambda_{{\rm eq}})}{\Phi
(\Lambda(t),\Lambda_{{\rm eq}})}}},
\end{equation}
with
\begin{equation}
\Phi(\Lambda,\Lambda_{{\rm eq}})=\exp\left(\int_1^{\Lambda}
{\rm d}\Lambda^{\prime}\,\alpha(\Lambda^{\prime},\Lambda_{{\rm eq}})\right).
\end{equation}
Finally, using eqs.~(\ref{dcts}) and the definitions of the $Y$-functions,
we obtain the scaling predictions
\begin{equation}
\frac{c(t,s)}{c_{{\rm pl}}(s)}\approx\frac{r^{\pm}(t,s)}{r_{{\rm pl}}^{\pm
}(s)}\approx\frac{\partial c(t,s)/\partial s}{\left(\partial c/\partial
s\right)_{{\rm pl}}(s)}\approx\frac{\Lambda(t)}{\Lambda(s)}\frac{\Phi
(\Lambda(s),\Lambda_{{\rm eq}})}{\Phi(\Lambda(t),\Lambda_{{\rm eq}})}
\label{eqscaling}
\end{equation}
throughout the non-equilibrium $\alpha$-regime,
which constitute the main result of this section.
The plateau values $c_{{\rm pl}}(s)$, $r_{{\rm pl}}^{\pm}(s)$,
and $\left(\partial c/\partial s\right)_{{\rm pl}}(s)$ in eq.~(\ref{eqscaling})
are the initial conditions for the dynamical
equations~(\ref{evolt}, \ref{evoll}).
These plateau values generalize the results~(\ref{eqplateau})
to the generic non-equilibrium situation at low temperature.
They are attained when the fast $(\beta)$ modes are extinct,
i.e., for $1\ll\theta=t-s\ll s$.
These quantities, which are non-trivial in general, remain to be determined,
since their evaluation does not pertain to the analysis
of the $\alpha$-regime stricto sensu.

In the case of the correlation function $c(t,s)$, the plateau value
\begin{equation}
c_{{\rm pl}}(s)=\Lambda(s)-1
\label{cneqplat}
\end{equation}
is simply obtained by replacing $\Lambda_{{\rm eq}}$ by $\Lambda(s)$ in the
expression~(\ref{eqplateau}) at equilibrium.
More generally, we have
\begin{equation}
c(t,s)\approx c_{\rm eq}(t-s;\Lambda(s))
\approx\Lambda(s)-1+{\rm e}^{-(t-s)/\Lambda(s)}
\label{cinst}
\end{equation}
[see eq.~\ref{ceq})] in the non-equilibrium $\beta$-regime.
This expression amounts to saying that the system is somehow at an
instantaneous equilibrium described by $\Lambda(s)$.
This simple description is only justified for moderate times ($t-s\ll s)$,
and for quantities, such as $c(t,s)$, which remain of order unity
for long waiting times and/or at low temperature.

Figure~1 illustrates the validity of the above description.
The zero-temperature correlation function $c(t,s)$ is plotted
against $\ln(t-s)$,
for different values of the waiting time~$s$, indicated on the curves.
The full lines show the result of a direct numerical integration
of eqs.~(\ref{eqfk}, \ref{eqgamk}).
The dashed lines show (to the left, in the $\beta$-regime) eq.~(\ref{cinst})
and (to the right, in the $\alpha$-regime) the scaling law~(\ref{eqscaling}),
with the plateau value~(\ref{cneqplat}).
The quantitative agreement between the exact numerical data and their
analytical description is increasingly convincing
as the waiting time $s$ increases.

The evaluation of the other plateau values,
namely $(\partial c/\partial s)_{{\rm pl}}(s)$ and $r_{{\rm pl}}^{\pm}(s)$,
requires more care, since these
quantities are already exponentially small at equilibrium.
A sketch of the method used for their derivation is given in Appendix~A.

\vskip 8.5cm{\hskip 0.8cm}
\includegraphics{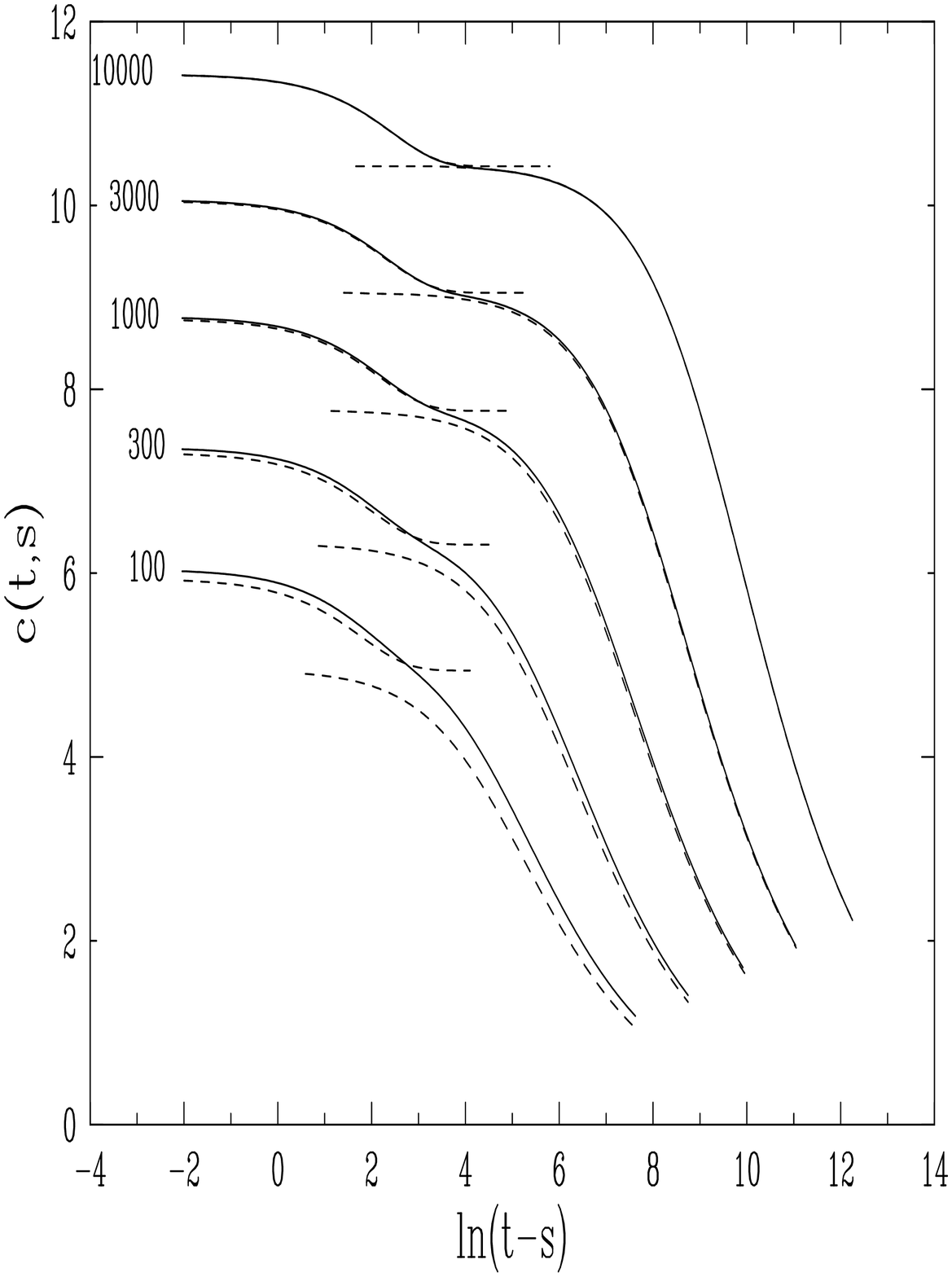}

\noindent {\bf Figure~1:}
Plot of the zero-temperature correlation function $c(t,s)$ against $\ln(t-s)$,
for different values of the waiting time~$s$, indicated on the curves.
Full lines: exact numerical results.
Dashed lines: left ($\beta$-regime): eq.~(\ref{cinst});
right ($\alpha$-regime): scaling law~(\ref{eqscaling}),
with plateau value~(\ref{cneqplat}).

\vskip .2cm

First, one finds
\begin{equation}
\left({{\frac{\partial c}{\partial s}}}\right)_{{\rm pl}}(s)\approx
\frac{\Lambda(s)^{2}{\rm e}^{-\Lambda(s)}}{\Lambda(s)^{2}{\rm e}^{-\Lambda
(s)}I(\Lambda(s))+1-\Lambda(s)}=\frac{\Lambda(s)-1}{t_{{\rm eq}}
\left[\Lambda(s)\right]},
\label{dcplat}
\end{equation}
where $t_{{\rm eq}}\left[\Lambda(s)\right]$ is obtained from
eq.~(\ref{teq}) by replacing $\Lambda_{{\rm eq}}$ by $\Lambda(s)$.
Though the expression~(\ref{dcplat}) is simple, since it can be obtained by
replacing $\Lambda_{{\rm eq}}$ by $\Lambda(s)$ in the equilibrium
result~(\ref{eqplateau}), its derivation is not.

The values of $r_{{\rm pl}}^{\pm}(s)$, obtained
along the same lines, have the more intricate expressions
\begin{eqnarray}
r_{{\rm pl}}^{+}(s) &\approx &\frac{\Lambda(s)^{2}{\rm e}^{-\Lambda
(s)}-\Lambda(s)A(\Lambda(s),\Lambda_{{\rm eq}})\left((\Lambda(s)-1)
{\rm e}^{-\Lambda(s)}I(\Lambda(s))-1+{{\frad{1}{\Lambda(s)^{2}}}}\right)}
{\Lambda(s)^{2}{\rm e}^{-\Lambda(s)}I(\Lambda(s))+1-\Lambda(s)}{,}\nonumber\\
r_{{\rm pl}}^{-}(s) &\approx &\frac{\Lambda(s)^{2}{\rm e}^{-\Lambda
(s)}-\Lambda(s)A(\Lambda(s),\Lambda_{{\rm eq}})\left(\Lambda(s)
{\rm e}^{-\Lambda(s)}I(\Lambda(s))-1-{{\frad{1}{\Lambda(s)}}}
+{{\frad{1}{\Lambda
(s)^{2}}}}\right)}{\Lambda(s)^{2}{\rm e}^{-\Lambda(s)}I(\Lambda
(s))+1-\Lambda(s)}{.}\nonumber\\
\label{rpmplat}
\end{eqnarray}

The scaling result~(\ref{eqscaling}) implies that the
\fd ratios $X^{\pm}(t,s)$ stay constant and equal to their plateau values
\begin{equation}
X_{{\rm pl}}^{\pm}(s)=\frac{r_{{\rm pl}}^{\pm}(s)}{\left(\partial
c/\partial s\right)_{{\rm pl}}(s)}
\end{equation}
throughout the non-equilibrium $\alpha$-regime, i.e., for $t-s\gg 1$.
Eqs.~(\ref{dcplat}, \ref{rpmplat}) lead to
\begin{eqnarray}
X_{{\rm pl}}^{+}(s) &\approx &1-\frac{{\rm e}^{\Lambda(s)}A(\Lambda
(s),\Lambda_{{\rm eq}})}{\Lambda(s)}\left((\Lambda(s)-1)
{\rm e}^{-\Lambda(s)}I(\Lambda(s))-1+{{\frac{1}{\Lambda(s)^{2}}}}\right),
\nonumber\\
X_{{\rm pl}}^{-}(s) &\approx &1-\frac{{\rm e}^{\Lambda(s)}A(\Lambda
(s),\Lambda_{{\rm eq}})}{\Lambda(s)}\left(\Lambda(s){\rm e}^{-\Lambda
(s)}I(\Lambda(s))-1-{{\frac{1}{\Lambda(s)}}}+{{\frac{1}{\Lambda(s)^{2}}}}
\right){.}
\end{eqnarray}
These expressions interpolate between the equilibrium values $X_{{\rm pl}}^{\pm
}(s)=1$ for $s\gg t_{{\rm eq}}$, and the non-trivial values
\begin{eqnarray}
X_{{\rm pl}}^{+}(s) &\approx &\frac{1}{\Lambda(s)}+\left(1-{{\frac{1}
{\Lambda(s)^{2}}}}\right)\frac{{\rm e}^{\Lambda(s)}}{\Lambda(s)I(\Lambda
(s))}\approx 1-\frac{2}{\Lambda(s)^{2}}-\frac{2}{\Lambda(s)^{3}}
-\frac{12}{\Lambda(s)^{4}}+\cdots{,}
\nonumber\\
X_{{\rm pl}}^{-}(s) &\approx &\left(1+{{\frac{1}{\Lambda(s)}}}-{{\frac{1}
{\Lambda(s)^{2}}}}\right)\frac{{\rm e}^{\Lambda(s)}}{\Lambda(s)I(\Lambda
(s))}\approx 1-\frac{3}{\Lambda(s)^{2}}-\frac{3}{\Lambda(s)^{3}}
-\frac{15}{\Lambda(s)^{4}}+\cdots\nonumber\\
\label{xplzero}
\end{eqnarray}
at zero temperature, and more generally in the aging regime, i.e.,
for $s\ll t_{{\rm eq}}$.

Figure~2 illustrates the above predictions.
The zero-temperature plateau values $X_{{\rm pl}}^\pm(s)$
are plotted against $\ln s$, for $s$ up to $10^5$.
The full lines show the limit values $\lim_{t\to\infty}X^\pm(t,s)$,
obtained by extrapolating the numerical solutions
of eqs.~(\ref{eqfk}, \ref{eqgamk}, \ref{eqhk}, \ref{eqdzk}).
The dashed lines show the analytical predictions~(\ref{xplzero}),
whose accuracy becomes extremely high for the larger values of
the waiting time $s$.

The above results
can be compared with the expression for the zero-temperature plateau value
of the \fd ratio associated with energy fluctuations,
derived in ref.~\cite{gl97}:
\begin{equation}
X_{{\rm pl}}^{{\rm energy}}(s)\approx\frac{\Lambda(s)(\Lambda(s)-1)}
{\Lambda(s)^{2}{\rm e}^{-\Lambda(s)}I(\Lambda(s))+(\Lambda(s)-1)^{2}}
\approx 1-\frac{2}{\Lambda(s)^{2}}-\frac{4}{\Lambda(s)^{3}}
-\frac{6}{\Lambda(s)^{4}}+\cdots
\label{xplener}
\end{equation}
The expressions~(\ref{xplzero}, \ref{xplener}) of the three \fd ratios
are very similar.
All these ratios asymptotically converge to the value 1,
characteristic of equilibrium, albeit with large correction terms,
which only decay as $1/\Lambda(s)^2\approx 1/(\ln s)^2$,
i.e., logarithmically slow with the waiting time~$s$.
More surprisingly, in the present case of density fluctuations, the difference
\begin{equation}
X_{\rm pl}^+(s)-X_{\rm pl}^-(s)\approx\frac{1}{\Lambda(s)}
\left(1-\frac{{\rm e}^{\Lambda(s)}}{\Lambda(s)I(\Lambda(s))}\right)
\approx\frac{1}{\Lambda(s)^2}+\frac{1}{\Lambda(s)^3}
+\frac{3}{\Lambda(s)^4}+\cdots,
\label{xdif}
\end{equation}
which is entirely due to the discontinuous slope of the Metropolis
acceptance rate~[see eq.~(\ref{metrodis})],
and hence dependent on our choice of a microscopic dynamics for the model,
exhibits the very same logarithmically slow fall-off in $1/(\ln s)^2$.

\vskip 8.5cm{\hskip 0.8cm}
\includegraphics{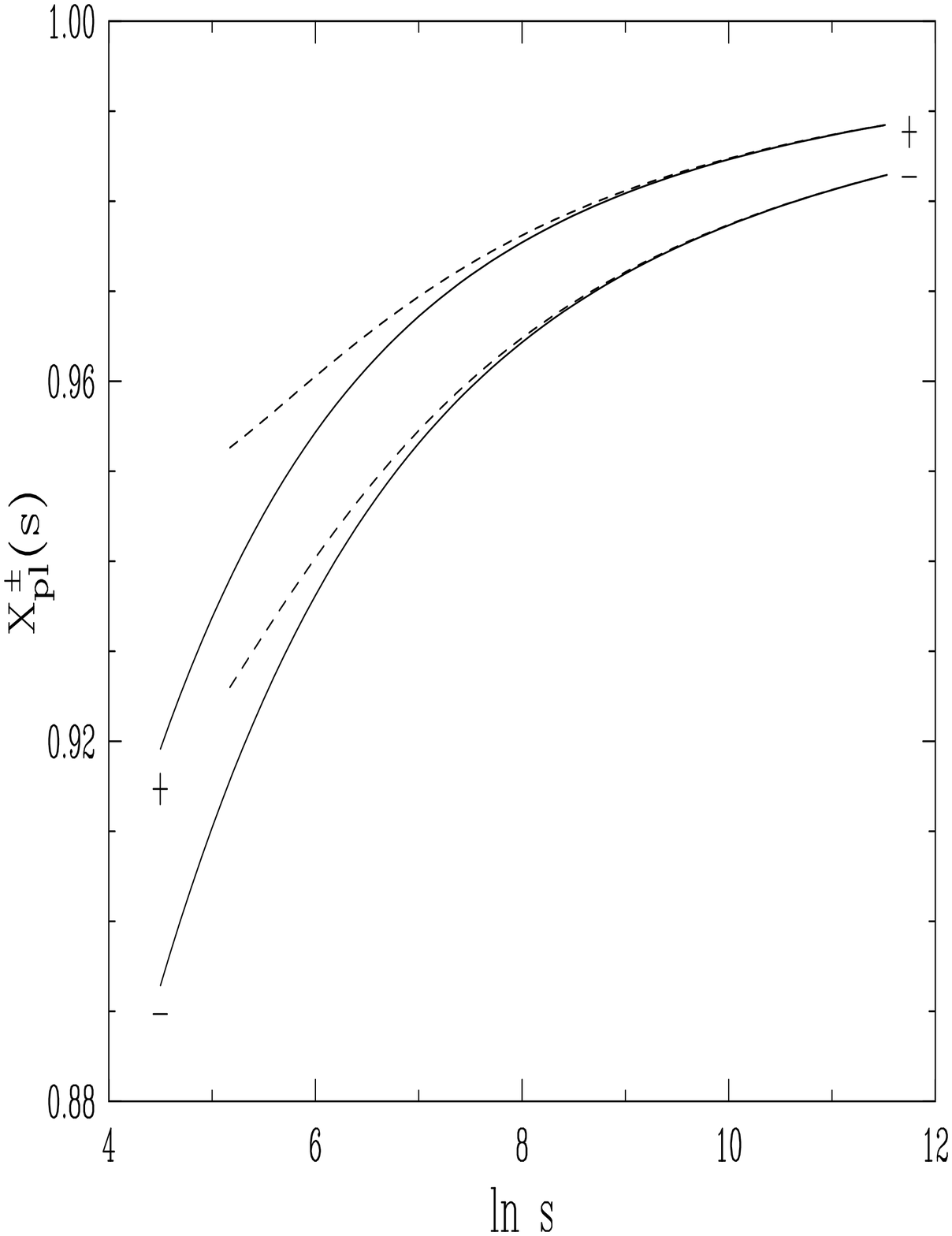}

\noindent {\bf Figure~2:}
Plot of the zero-temperature plateau values
$X_{\rm pl}^+(s)$ (upper data) and $X_{\rm pl}^-(s)$ (lower data),
against $\ln s$.
Full lines: exact numerical results.
Dashed lines: analytical predictions~(\ref{xplzero}).

\vskip .2cm

To close up, we give a more detailed account
of the dependence on both time scales $s$ and $t$
of the scaling law~(\ref{eqscaling}):

\begin{itemize}
\item{
At zero temperature, and more generally in the aging regime
$(s\ll t_{{\rm eq}})$, we have
\begin{equation}
\alpha(\Lambda,\infty)=\frac{\Lambda ^{2}{\rm e}^{-\Lambda}I(\Lambda)
+1}{\Lambda\left(\Lambda ^{2}{\rm e}^{-\Lambda}I(\Lambda)+1-\Lambda
\right)}\approx\frac{1}{2}+\frac{1}{2\Lambda}
-\frac{1}{\Lambda^2}-\frac{7}{2\Lambda^3}+\cdots,
\end{equation}
hence
\begin{eqnarray}
\Phi(\Lambda(s),\infty)
&\approx&\left(\Lambda(s){\rm e}^{\Lambda(s)}\right)^{1/2}
\left(1+\frac{1}{\Lambda(s)}+\frac{9}{4\Lambda(s)^2}+\cdots\right)\nonumber\\
&\approx&s^{1/2}\ln s
\left(1+\frac{\ln\ln s}{\ln s}+\frac{4\ln\ln s-9}{4(\ln s)^2}+\cdots\right).
\end{eqnarray}
Therefore, in the aging regime $(s,\,t\ll t_{{\rm eq}})$,
the right-hand-side of the scaling law~(\ref{eqscaling}) roughly obeys a
square-root behavior, namely
\begin{equation}
\frac{\Lambda(t)}{\Lambda(s)}\frac{\Phi(\Lambda(s),\infty)}
{\Phi(\Lambda(t),\infty)}\approx\left({{\frac{s}{t}}}\right)^{1/2}
\left(1+\frac{1}{4}\left(\frac{1}{(\ln t)^2}-\frac{1}{(\ln s)^2}\right)
+\cdots\right).
\label{isqrt}
\end{equation}
}
\item{
In the opposite limit $(s\gg t_{{\rm eq}})$,
describing the convergence towards equilibrium,
the scaling function $\Phi$ blows up exponentially, as
\begin{equation}
\Phi(\Lambda(s),\Lambda_{{\rm eq}})
\approx C(\Lambda_{{\rm eq}}){\rm e}^{s/t_{{\rm eq}}},
\end{equation}
with an $s$-independent prefactor.
In this regime, the scaling law~(\ref{eqscaling})
therefore exhibits an exponential decay,
\begin{equation}
{{\frac{\Lambda(t)}{\Lambda(s)}\frac{\Phi(\Lambda(s),\Lambda_{{\rm eq}})}
{\Phi(\Lambda(t),\Lambda_{{\rm eq}})}}}\approx{\rm e}^{-(t-s)/t_{{\rm eq}}},
\end{equation}
characteristic of equilibrium properties.
}
\end{itemize}

\section{Summary and discussion}

The present work, a sequel of refs.~\cite{gbm,gl96,gl97},
is devoted to a thorough study of the density correlation
and response functions
and of the associated \fd ratios for the Backgammon model,
a mean-field dynamical model introduced in the context of glassy
dynamics~\cite{ritort}.
We thus provide analytical predictions for the most
salient findings of Franz and Ritort~\cite{fr3} for the relaxational
dynamics of the density fluctuations of the model,
both at equilibrium and away from equilibrium.

Let us first summarize our main results.
Consider the situation at equilibrium and low temperature.
The correlations in the Backgammon model possesses a spectrum of relaxation
times with an exponential separation of time scales,
between a $\beta$-regime of fast relaxation,
involving the microscopic time scale, and a slow
$\alpha$-regime, characterized by the relaxation time
$t_{{\rm eq}}\approx 2{\rm e}^{\Lambda_{{\rm eq}}}/\Lambda_{{\rm eq}}
\approx 2{\rm e}^{\beta}/\beta^{2}$,
where $\Lambda_{{\rm eq}}\approx\beta$ is the fugacity of the system.
The two-time correlation and response functions are stationary,
i.e., they only depend on the time difference $\theta=t-s$.
The density correlation function reads~[see eq.~(\ref{ceq})]
\begin{equation}
c_{{\rm eq}}(\theta)\approx\underbrace{{\rm e}^{-\theta /\Lambda_{{\rm eq}%
}}}_{\displaystyle{\beta}\hbox{-regime}}+\underbrace{\left(\Lambda_{{\rm %
eq}}-1\right) {\rm e}^{-\theta /t_{{\rm eq}}}}_{\displaystyle{\alpha}%
\hbox{-regime}}.
\label{ceqdis}
\end{equation}
This expression accounts for the superposition of the fast $\beta$-relaxation
($\theta\sim 1$ or $\Lambda_{{\rm eq}}$), and the slow $\alpha$-relaxation
($\theta\sim t_{{\rm eq}}$).
As a consequence, eq.~(\ref
{ceqdis}) exhibits a well-defined plateau value~[see eq.~(\ref{eqplateau})]
\begin{equation}
(c_{{\rm eq}})_{{\rm pl}}\approx\Lambda_{{\rm eq}}-1
\end{equation}
throughout the crossover between both regimes, i.e., for $1\ll\theta\ll t_{%
{\rm eq}}$.
The same conclusions hold for the response, with
\begin{equation}
r_{{\rm eq}}(\theta)=-\frac{{\rm d}}{{\rm d}\theta}c_{{\rm eq}}(\theta),
\end{equation}
which expresses the \fd theorem.

One of the most striking outcomes of refs.~\cite{gl96,gl97}, further
extended in the present work, is that the exponential separation of fast
($\beta$) and slow ($\alpha$) modes still holds in a generic
non-equilibrium situation.
This simplifying feature allows an analytical
treatment of the non-equilibrium $\alpha$-regime at low temperature.
Throughout this regime, we predict that the density correlation and response
functions obey multiplicative scaling laws
of the form~[see eq.~(\ref{eqscaling})]
\begin{equation}
\frac{c(t,s)}{c_{{\rm pl}}(s)}\approx\frac{r^{\pm}(t,s)}{r_{{\rm pl}}^{\pm
}(s)}\approx\frac{\partial c(t,s)/\partial s}{\left(\partial c/\partial
s\right)_{{\rm pl}}(s)}\approx\frac{\Lambda(t)}{\Lambda(s)}\frac{\Phi
(\Lambda(s),\Lambda_{{\rm eq}})}{\Phi(\Lambda(t),\Lambda_{{\rm eq}})}.
\label{eqscadis}
\end{equation}

These scaling laws hold at low temperature ($t_{{\rm eq}}\gg 1$) and
throughout the $\alpha$-regime ($s\gg 1$, $t-s\gg 1$), irrespective of the
ratios $s/t_{{\rm eq}}$ or $(t-s)/t_{{\rm eq}}$.
When these ratios are
small, i.e., at zero temperature, and more generally in the aging regime,
the rightmost side of eq.~(\ref{eqscadis}) assumes an approximate
inverse-square-root law $(s/t)^{1/2}$~[see eq.~(\ref{isqrt})].
Conversely, when the above ratios are
large, i.e., in the regime of convergence towards equilibrium at a low but
finite temperature, this rightmost side exhibits an exponential fall-off, of
the form $\exp(-(t-s)/t_{\rm eq})$, characteristic of the equilibrium
$\alpha$-relaxation.

Moreover, in the case of the correlation function, the crossover between
the non-equilibrium $\beta$- and $\alpha$-regimes can be described by a
simple additive formula, generalizing the equilibrium result~(\ref{ceqdis}), as
\begin{equation}
c(t,s)\approx\underbrace{{\rm e}^{-(t-s)/\Lambda(s)}}_{\displaystyle{\beta
}\hbox{-regime}}+\underbrace{\left(\Lambda(s)-1\right)\frac{\Lambda(t)}{%
\Lambda(s)}\frac{\Phi(\Lambda(s),\Lambda_{{\rm eq}})}{\Phi(\Lambda
(t),\Lambda_{{\rm eq}})}}_{\displaystyle{\alpha}\hbox{-regime}}.
\label{cdis}
\end{equation}
This result, illustrated in Figure~1, amounts to saying that in the
non-equilibrium $\beta$-regime the system is somehow at an instantaneous
equilibrium, described by $\Lambda(s)$.
The validity of such a description
is related to the simplicity of the non-equilibrium
plateau value~[see eq.~(\ref{cneqplat})]
\begin{equation}
c_{{\rm pl}}(s)=\Lambda(s)-1.
\end{equation}
The same holds for the derivative $\partial c(t,s)/\partial s$.
On the contrary, for the response functions $r^{\pm}(t,s)$,
there is still a well-defined separation of a fast
and a slow component away from equilibrium,
but this separation is non-trivial,
as it involves the plateau values $r_{\rm pl}^\pm(s)$,
which are complicated functions of $s$,
even at zero temperature~[see eqs.~(\ref{rpmplat})].

It is worth comparing the results of this paper for the correlation and
response of the density with those found for the correlation and response
of the energy~\cite{gl96,gl97}.
First, as underlined in ref.~\cite{fr3},
the density correlation has an appreciable variation over short time scales,
i.e., in the $\beta $-regime, while the energy correlation remains
essentially constant in this regime.
Second, in the $\alpha$-regime, the scaling law~(\ref{eqscadis})
of the correlation and response functions of the density
is very similar to that of the energy~\cite{gl97}.
As it turns out, both scaling functions only differ by the prefactor
$\Lambda(t)/\Lambda(s)$.
Finally, in the present case of density fluctuations,
because of the choice of a Metropolis dynamics,
it is necessary to introduce two response functions, $r^\pm(t,s)$,
and accordingly two \fd ratios, $X^\pm(t,s)$.
This subtlety was not needed in the case of energy fluctuations.
A surprising outcome of the present study is that the difference
$X_{{\rm pl}}^{+}(s)-X_{{\rm pl}}^{-}(s)$,
which reflects the choice of the microscopic dynamics of the model,
persists in the $\alpha$-regime,
where it exhibits a logarithmically slow fall-off~[see eq.~(\ref{xdif})].

As can be seen on these results, the Backgammon model is helpful in giving
clear-cut analytical answers to some recurrent questions in the ongoing
investigation of the slow dynamics of non-equilibrium systems.
This knowledge in turn is of interest to see whether
one can discriminate between non-equilibrium systems by
their relaxational behavior, encoded in the correlation and response, or
equivalently in the behavior of the fluc\-tu\-a\-tion-dis\-si\-pa\-tion ratio.
Thus, in the present case, is the Backgammon model more akin to a
coarsening system, or to a mean-field spin-glass model, or yet to another case?

In some respects, the quantity $\Lambda (t)$, which is omnipresent in our
description of the dynamics of the model, plays the role of the mean domain
size $L(t)$, which is the only characteristic length scale in a system
undergoing phase ordering~\cite{langer,bray}.
Another common salient feature
of the present model and of coarsening systems is that the scaling
form~(\ref {eqscadis}) of the correlation function $c(t,s)$ involves the ratio
of a function of $\Lambda (t)$ to the same function of $\Lambda (s)$.
On the other hand, as far as the fluc\-tu\-a\-tion-dis\-si\-pa\-tion ratio is
concerned, there is a maximal discrepancy between the present model
at zero temperature,
where both fluc\-tu\-a\-tion-dis\-si\-pa\-tion ratios $X^{\pm }(t,s)$ go
asymptotically to unity (with a large logarithmic correction, illustrated in
Figure~2), and coarsening systems, for which the
fluc\-tu\-a\-tion-dis\-si\-pa\-tion ratio is observed
to go to zero~\cite{ckp,barrat,berthier}.
(The interpretation of the latter behavior is that
the long-time response comes from the movement of the domain walls,
and therefore becomes more and more negligible as the
domains grow~\cite{barrat,berthier}.)
This discrepancy is actually due to the fact that
the critical temperature of the Backgammon model is $T_c=0$.
As a consequence, the low-temperature phase, where aging persists forever,
is reduced to the critical point of the model.
More generally, as will be presented elsewhere~\cite{ustocome},
the fluc\-tu\-a\-tion-dis\-si\-pa\-tion ratio
exhibits special features for systems quenched to their critical point.

\subsubsection*{Acknowledgements}

We wish to thank S.~Franz and F.~Ritort
for interesting discussions during the elaboration of this work.

\newpage
\appendix
\section{Derivation of non-equilibrium plateau values}

In this Appendix, we present the method used in subsection 6.2
for the evaluation of the plateau values
$r_{{\rm pl}}^{\pm}(s)$ and $(\partial c/\partial s)_{{\rm pl}}(s)$
in the generic non-equilibrium situation.

Let us take the example of the latter quantity.
First, eq.~(\ref{dcts}) implies
\begin{equation}
\left({{\frac{\partial c}{\partial s}}}\right)_{{\rm pl}}(s)
=-\Lambda(s)(Y_{\zeta})_{{\rm pl}}(s).
\end{equation}
Second, $(Y_{\zeta})_{{\rm pl}}(s)$ can be estimated by
performing a Laplace transform on eqs.~(\ref{ydzeta}, \ref{eqdzeta0}).
We thus obtain after some algebra
\begin{equation}
\left({{\frac{\partial c}{\partial s}}}\right)_{{\rm pl}}(s)\approx
-\Lambda(s)\frac{\Lambda(s)S_{\zeta}(s)+(\Lambda(s)-1)f_{1}(s)}{\Lambda
(s)^{2}{\rm e}^{-\Lambda(s)}I(\Lambda(s))+1-\Lambda(s)},
\end{equation}
with
\begin{equation}
S_{\zeta}(s)=\Lambda(s)\int_{0}^{1}{\rm d}x\,\frac{{\rm e}^{-\Lambda(s)x}}
{1-x}Z(x,s,s),
\label{sz}
\end{equation}
where $Z(x,s,s)$ is given in eq.~(\ref{zss}) in terms of the generating
series $F(x,s)$ of the occupation probabilities.

Generalizing the equilibrium results derived in Sec.~5,
we anticipate that $S_{\zeta}(s)$ is exponentially small in $\Lambda(s)$.
As a consequence, it is not legitimate to estimate $F(x,s)$
by just replacing $\Lambda_{{\rm eq}}$ by $\Lambda(s)$ in the equilibrium
result~(\ref{feqx}).
In other words, we first have to derive a better evaluation of the
occupation probabilities in the $\alpha$-regime at low temperature.
This can be done by seeking a
solution to the first partial differential equation of eq.~(\ref{partialad})
in the form
\begin{equation}
F(x,s)={{\frac{\Lambda(s)-1+{\rm e}^{(x-1)\Lambda(s)}}{\Lambda(s)}}}
+{\rm e}^{(x-1)\Lambda(s)}\Delta(x,s),
\label{fxs}
\end{equation}
where the first term represents the occupation probabilities at the
instantaneous equilibrium described by $\Lambda(s)$~[see eq.~(\ref{feqx})],
while the correction term $\Delta(x,s)$ is assumed to be small.
Eq.~(\ref{partialad}) yields a differential equation for $\Delta(x,s)$,
which can be integrated as
\begin{equation}
\Delta(x,s)=-\frac{{\rm d}\Lambda(s)}{{\rm d}s}\int_{x}^{1}{\rm d}y\,
\frac{{\rm e}^{(1-y)\Lambda(s)}-(1-y)\Lambda(s)-1}{(1-y)\Lambda(s)}.
\label{deltaxs}
\end{equation}
This expression is proportional to ${\rm d}\Lambda(s)/{\rm d}s$,
i.e., exponentially small in $\Lambda(s)$, as expected.
The result~(\ref{deltaxs}) has several consequences.
\begin{itemize}
\item{
Setting $x=0$ in eq.~(\ref{deltaxs}),
we obtain $\Delta(0,s)\approx -({\rm d}\Lambda(s)/{\rm d}s)
I(\Lambda(s))/\Lambda(s)$.
Remarkably enough, we thus recover the
evolution equation~(\ref{dlambda}) for $\Lambda(s)$ itself.
}
\item{
By expanding eq.~(\ref{fxs}) to second order in $(x-1)$, we obtain
\begin{equation}
c(s,s)=\langle N_{1}(s)^{2}\rangle-1=\sum_{k\ge 0}k^{2}\,f_{k}(s)-1=\left.
\frac{{\partial}^{2}F(x,s)}{{\partial}x^{2}}\right|_{x=1}
\approx\Lambda(s)-\frac{\Lambda(s)A(\Lambda(s),\Lambda_{{\rm eq}})}{2}.
\end{equation}
The second moment of the density fluctuations is thus given by
the simple formula corresponding to the Poisson law of parameter $\Lambda(s)$,
namely $c(s,s)=\langle N_{1}(s)^{2}\rangle-1=\Lambda(s)$,
based on the picture of an instantaneous equilibrium described by $\Lambda(s)$,
up to an exponentially small correction, proportional to
${\rm d}\Lambda/{\rm d}s\approx A(\Lambda(s),\Lambda_{{\rm eq}})$,
}
\item{
Similarly, we can evaluate the initial values of the \fd ratios $X^{\pm}(s,s)$.
We have indeed $r^{\pm}(s,s)\approx\partial c(s,s)/\partial
s\approx 1/\Lambda(s)$, and
\begin{eqnarray}
\frac{\partial c(s,s)}{\partial s}-r^{\pm}(s,s) &=&1-{\rm e}^{-\beta
}f_{0}(s)+\frac{1}{\Lambda(s)}(f_{1}(s)-c(s,s))\nonumber\\
&\approx &\left({{\frac{1}{2}}}+{{\frac{1}{\Lambda(s)^{2}}}}\right)
A(\Lambda(s),\Lambda_{{\rm eq}}),
\end{eqnarray}
hence
\begin{equation}
X^{\pm}(s,s)\approx 1-\left({{\frac{1}{2}}}+{{\frac{1}{\Lambda(s)^{2}}}}
\right)\Lambda(s)A(\Lambda(s),\Lambda_{{\rm eq}}).
\end{equation}
The violation of the \fd theorem at initial times is thus exponentially small,
as it is again proportional to ${\rm d}\Lambda(s)/{\rm d}s$.
We have in particular
\begin{equation}
X^{\pm}(s,s)\approx 1-\left({{\frac{1}{2}}}+{{\frac{1}{\Lambda(s)^{2}}}}
\right)\frac{1}{I(\Lambda(s))}
\approx 1-\frac{\Lambda(s){\rm e}^{-\Lambda(s)}}{2}
\end{equation}
at zero temperature.
}
\item{
Finally, coming back to our main goal,
i.e., the derivation of the plateau value
of $(\partial c/\partial s)_{{\rm pl}}(s)$,
we insert eqs.~(\ref{fxs}, \ref{deltaxs})
into the expression~(\ref{zss}) for $Z(x,s,s)$,
and then perform the integral~(\ref{sz}).
All the integrals brought by the correction term $\Delta(x,s)$
can be expressed in terms of $I(\Lambda(s))$ and of elementary functions.
We are thus left with
\begin{equation}
\left({{\frac{\partial c}{\partial s}}}\right)_{{\rm pl}}(s)\approx
\frac{\Lambda(s)^{2}{\rm e}^{-\Lambda(s)}}{\Lambda(s)^{2}{\rm e}^{-\Lambda
(s)}I(\Lambda(s))+1-\Lambda(s)},
\end{equation}
i.e., the result announced in eq.~(\ref{dcplat}).
The results~(\ref{rpmplat}) have been derived along the very same lines.
}
\end{itemize}

\end{document}